\documentclass[aps,pre,
showkeys,
longbibliography,
superscriptaddress,
reprint,                
twocolumn,              
nofootinbib,
floatfix]{revtex4-2}

\usepackage[T1]{fontenc}
\usepackage[utf8]{inputenc}
\usepackage{amsmath}
\usepackage{amsfonts}
\usepackage{graphicx}
\usepackage{xcolor}

\DeclareMathOperator{\sign}{\mathrm{sgn}}
\DeclareMathOperator{\Prob}{\mathrm{Prob}}

\usepackage[colorlinks=true,hyperfootnotes=true,breaklinks=true,allcolors=teal]{hyperref}
\usepackage[capitalise]{cleveref}

\begin{document}

\title{Extending the Biswas--Chatterjee--Sen model with anticonformists and inflexibles}

\author{Amit Pradhan}
\thanks{\href{https://orcid.org/0009-0007-0442-5443}{0009-0007-0442-5443}}
\email{prdhanamit84@gmail.com}
\affiliation{\href{https://ror.org/01e7v7w47}{University of Calcutta}, Department of Physics, 92 Acharya Prafulla Chandra Road, Kolkata, 700009, India}

\author{Parongama Sen}
\thanks{\href{https://orcid.org/0000-0002-4641-022X}{0000-0002-4641-022X}}
\email{psphy@caluniv.ac.in}
\affiliation{\href{https://ror.org/01e7v7w47}{University of Calcutta}, Department of Physics, 92 Acharya Prafulla Chandra Road, Kolkata, 700009, India}

\author{Krzysztof Malarz}
\thanks{\href{https://orcid.org/0000-0001-9980-0363}{0000-0001-9980-0363}}
\email{malarz@agh.edu.pl}
\affiliation{\href{https://ror.org/00bas1c41}{AGH University}, Faculty of Physics and Applied Computer Science, al.~Mickiewicza~30, 30-059 Krak\'ow, Poland}

\begin{abstract}
Originally, the Biswas--Chatterjee--Sen model was shown to exhibit an order/disorder phase transition for a sufficiently large number of negative interactions among actors. In this paper, the model is extended by the existence of anticonformists and inflexibles. Anticonformists are actors who define themselves in opposition to the group and may intentionally reject what most people accept, while inflexibles are those who do not change their opinions at all. Both discrete and continuous opinions are considered. With direct Monte Carlo simulations and mean-field calculations, we check the influence of fractions of anticonformists and inflexibles on the mean opinion in the system. With the mean-field calculations, we identify ranges of fractions of anticonformists where an ordered phase of the system is available. The results of the mean-field calculations perfectly match the results of the Monte Carlo simulations. We consider inflexibles adhered: (i) to extreme opinions; (ii) to specific opinions, and (iii) chosen independently of their initial opinion. For inflexibles adhered to specific and extreme opinions, they play a role of an effective bias suppressing the disordered phase in the system. The qualitative results of introducing anticonformists (inflexibles) in various ways (discrete/continuous opinions and annealed/quenched disorder) are roughly the same. However, for the model extended by inflexibles, we can observe a systematic shift of the mean order parameter to its higher values for quenched disorder compared with annealed disorder. On the other hand, for  anticonformists modeled with a continuous space of opinions, we can observe a systematic shift of the mean order parameter to its higher values compared with the discrete space of opinions. 
\end{abstract}

\date{September 15th, 2026}

\maketitle

\section{Introduction}

Sociophysics \cite{Sen_2014,Matjaz_2019,Jusup_2022,SI_70Galam_40Sociophysics} tackles the problems of social science with methods and tools of statistical physics.
Among these problems, the formation and dynamics of public opinion \cite{Stauffer2009,Castellano_2009,Sobkowicz_2019,Hassani_2022,Zachary_2022} is one of the core topics of sociophysics. 

The opinion formation models may be classified into discrete or continuous ones.
The latter are represented by the models of \citeauthor{Weisbuch_2002} \cite{Weisbuch_2002,Malarz2006b} or \citeauthor{Hegselmann-2002} \cite{Hegselmann-2002}. In such models, opinions are represented as real numbers, for example, taking values in the interval $[-1;+1]$.
Among discrete models one should mention:
voter \cite{Clifford_1973,Holley_1975,Liggett_1999,Liggett_2005}, 
majority-rule \cite{Galam_2002}, 
Sznajd \cite{Sznajd-2000} or models based on social impact \cite{Latane-1976,Latane-1981,Nowak-1990,Latane_Nowak_1994,1902.03454,2002.05451,2010.15736,2211.04183,Maslyk_2023,2405.05114}.
In such models, the opinions are represented as integers, in the simplest case with only two opinions in $\{-1,+1\}$.
However, for discrete models,  more than two opinion values (but still with their finite number) were also investigated for the voter model \cite{Hadzibeganovic_2008,Vazquez_2004,Szolnoki_2004,Castello_2006,Mobilia_2011,Starnini_2012,Mobilia_2023,2405.05114}, 
the Sznajd model \cite{Rodrigues_2005,Kulakowski2010,Doniec_2022,2405.05114}, 
the Latan\'e model \cite{2405.05114,Maslyk_2023,2211.04183,2002.05451,1902.03454},
the majority-rule model \cite{Gekle-2005,Lima_2012,Galam_2013,Wu_2018,Zubillaga_2022} 
and in other studies \cite{Vazquez_2003,Xiong_2017,Ozturk_2013,Martins_2020,Li_2022}.

The kinetic model of opinion proposed by Biswas, Chatterjee and Sen (BChS) \cite{Biswas_2012,Sen_2012} was originally formulated in both versions: continuous and discrete.
Soon after its introduction, the BChS model was followed by several modifications \cite{Crokidakis_2014a,Crokidakis_2014b} with the most recent presented in References~\onlinecite{Oliveira_2025,Vieira_2024,Oliveira_2024,Biswas_2024} (see also References~\onlinecite{Biswas_2023,Goswami_2015} for reviews).

In the BChS model, the interactions can take positive or negative values. 
In this paper, we enrich the BChS model with inflexibles and anticonformists.
Inflexibles are agents who never change their own opinion but can influence others, and anticonformists are those who go against the opinion of others.

In this context, it can be mentioned that negative interaction in the BChS model has been misinterpreted as the presence of contrarian behavior in a previous study \cite{Crokidakis_2014a}. However,  in the BChS model,
negative interactions are stochastically chosen during each interaction without identifying specific agents to have contrarian behavior in the sense that they have negative interactions all the time. In the  model considered in the present paper, the  presence of anticonformists (sometimes also called contrarians) provides an independent, additional source of noise, such agents necessarily go against the opinion of the agents they interact with.

Using both mean-field theory calculations and Monte Carlo simulation results, we show how the fraction of anticonformists and the fraction of inflexibles separately influence the original model results.
We consider both a discrete and a continuous spectrum of opinions.
In addition, various variants of the disorder (annealed or quenched) are considered.
For an annealed disorder, every actor decides to behave as anticonformists (or inflexibles) with a given probability corresponding to the assumed fraction of them in the society. 
For quenched disorder, the identification (and technically also position in the network) of inflexibles and anticonformists is fixed. In both cases, the fraction of the anticonformist (or inflexibles) acts as a  model control parameter.

Some ways of introduction of the inflexibles may shift the original curve of the order parameter versus noise level towards higher values of the order parameter and may remove the disordered phase altogether from the system.

In contrast, introducing anticonformists leads to a reduction of the order parameter.
For some range of their abundance, only a disordered phase is observed.
The results of the mean-field calculations perfectly match the results of direct Monte Carlo simulations.

The paper is organized as follows: \Cref{sec:model} introduces the model and necessary modifications to deal with groups of anticonformists and inflexibles; 
\Cref{sec:results} presents the results (including mean-field calculations in \Cref{sec:Mean-field_calculations,sec:Mean-field_calculations_for_anticonformists,sec:Mean-field_calculations_for_inflexibles} and Monte Carlo simulations in \Cref{sec:Monte_Carlo_simulations})
and \Cref{sec:discussion} is devoted to their discussion.

\section{\label{sec:model}Model}

In the original model formulation \cite{Biswas_2012} the time evolution of the opinion $o_i$ of the actor $i$ is given as
\begin{equation}
o_i(t+1)= o_i(t) + \mu_{ij}o_j(t),
\label{eq:model}
\end{equation}
where the elements of the matrix $\mu_{ij}$ (alignment index between agents $i$ and $j$) take random values from $[-1;+1]$ (in continuous version) or $\mu=\pm 1$ (in discrete version). 
The fraction $p$ of these values is negative and $p$ plays the role of the noise level.

In the continuous version of the model opinions $o_i\in[-1;+1]$, while in the discrete version only three values of the opinions are considered, $o_i\in\{-1,0,+1\}$.
Anyway, in both versions, if the anticipated value of the opinion according to \Cref{eq:model} is greater (less) than $+1$ ($-1$) then it is set to $+1$ ($-1$).
The latter leads to the rule
\begin{equation}
o_i(t+1)= \sign\big(o_i(t) + \mu_{ij}o_j(t)\big)
\label{eq:model-disc}
\end{equation}
in the discrete version of the model.

Originally, the order parameter of the model with $N$ actors in total is defined as
\begin{equation}
O(t) = \dfrac{1}{N}\left|\sum_i o_i(t)\right|.
\label{eq:order}
\end{equation}

The results of the  original BChS model \cite{Biswas_2012} can be summarized as follows: the mean-field (on a fully connected graph and for the discrete version of opinions) model shows a order-disorder phase transition as $p$ exceeds a critical value $p_c$. 
The critical value $p_c$ depends on whether the alignment index values $\mu_{ij}$ are randomly chosen from a discrete set ($-1$ or $+1$) or continuously and uniformly between $-1$ to $+1$.
The alignment index $\mu_{ij}$ can also be considered as a quenched or annealed variable that does not affect the results. The fully connected model was found to belong to the Ising mean-field universality class. Later, many variations of the original model had been studied \cite{Biswas_2023}.

\subsection{\label{sec:anticonformists}Anticonformists}

According to the Cambridge Dictionary \cite{CambridgeDictionary}, nonconformist is `someone who lives and thinks in a way that is different from other people'.
An anticonformist goes a step further: they often define themselves in opposition to the group and may intentionally reject what most people accept \cite{1906.06094}.

We define actor $i$ as an anticonformist setting $\mu_{ij}=-1$ when he/she interacts with any other actor denoted by $j$.
We set their fraction in society as $c$.
For the remaining part of the society [\textit{i.e.}, the fraction $(1-c)$ of actors] the alignment index $\mu_{ij}$ takes random values $\mu=\pm 1$ (in discrete version) or it is randomly drawn from $[-1,1]$ (in continuous versions) as in the original model formulation. 
In both cases (discrete/continuous opinions) the fraction $p$ of the values $\mu_{ij}$ in this fraction is negative.

\subsection{\label{sec:inflexibles}Inflexibles}

Inflexibles are actors who have fixed opinions and are unable (or unwilling) to change them.
For modeling the presence of inflexibles in the society, we assume several scenarios.
In the most extreme assumption: the fraction $z$ of the actors adheres to an extreme opinion (either $+1$ or $-1$) and never changes it.
We also consider the case where the fraction $z$ of inflexibles is mixed with the believers for both extreme opinions ($+1$ and $-1$).
Finally, also the scenario where actors with neutral opinions are inflexibles is considered.
The remaining (flexible) part of the population [\textit{i.e.}, the fraction $(1-z)$ of actors] follows the rule given in \Cref{eq:model}.

\section{\label{sec:results}Results}

\subsection{\label{sec:Mean-field_calculations}Mean-field calculations}

In the mean-field description of the kinetic exchange opinion model and discrete opinions, the state of the system is characterized by the fractions $f_+(t)$, $f_-(t)$, $f_0(t)$ denoting, respectively, the densities of agents holding opinions $+1$, $-1$ and $0$. These satisfy the normalized condition
\begin{equation}
    f_+(t)+f_-(t)+f_0(t) = 1.
\end{equation}
For mean-field calculations, it is convenient to introduce the order parameter 
\begin{equation}
    O(t) = f_+(t) - f_-(t)
\end{equation}
and the activity or the non-neutral fraction
 \begin{equation}
     s(t) = f_+(t) + f_-(t).
 \end{equation}
Thus, $s\in [0;1]$ measures how many agents hold a non-neutral opinion, while $O$ captures net polarization.

\subsection{\label{sec:Mean-field_calculations_for_anticonformists}Mean-field calculations for anticonformists}

The anticonformist version of the model is defined in \Cref{sec:anticonformists}. The mean-field transition rates between the three discrete opinion states are given by
\begin{equation}
\begin{split}
    \omega_{+1 \to 0} = \omega_{0 \to -1} = (1-c)[(1-p)f_-+pf_+] + cf_+,\\ 
    \omega_{-1 \to 0} = \omega_{0 \to +1} = (1-c)[(1-p)f_++pf_-] + cf_-,\\
    \omega_{+1 \to -1} = \omega_{-1 \to +1} = 0.
\end{split}
\end{equation}

Using these rates, the mean-field dynamics can be expressed directly in terms of the order parameter $O$ and activity $s$ as
\begin{equation}
\label{activity equation for anticonformist}
    \frac{ds}{dt} = s-\frac{3}{2}s^2 + \frac{O^2}{2}[(1-c)(1-2p)-c],
\end{equation}
\begin{equation}
\label{order parameter equation for anticonformist}
\begin{split}
    \frac{dO}{dt} = O\bigg[(1-s)\{(1-c)(1-2p)-c\}\\
    +\frac{s}{2}\{(1-c)(1-2p)-c-1\}\bigg].
\end{split}
\end{equation}
\begin{description}
\item[Disordered fixed point and its stability]
The disordered fixed point is obtained by setting $O= 0$ in \Cref{activity equation for anticonformist}, which gives the physically relevant disordered fixed point $(O^*,s^*) = (0,2/3)$. 

Linearizing the dynamical \Cref{activity equation for anticonformist,order parameter equation for anticonformist} around this point yields two eigenvalues:
\begin{equation}
\label{eigenvalue of jacobian}
    \lambda_s = -1, \quad \lambda_O = \frac{2}{3}\left[(1-c)(1-2p)-c\right]-\frac{1}{3}.
\end{equation}
Since $\lambda_s<0$, the stability of the disordered fixed point is controlled by $\lambda_O$. For $\lambda_O<0$, the disordered fixed point is stable, while for $\lambda_O>0$, it becomes a saddle. The latter condition corresponds to the emergence of the ordered fixed points.

The phase boundary is obtained from $\lambda_O=0$, leading to
\begin{equation}
\label{eq:pc_vs_c}
p_c(c) = \frac{1-4c}{4(1-c)}.
\end{equation}

This leads to two regimes:
\begin{description}
\item[$0 < c < 1/4$]
The critical noise $p_c(c)$ decreases with increasing $c$ and vanishes at $c = 1/4$. The system exhibits an order–disorder transition at $p = p_c(c)$, separating ordered $p < p_c(c)$ and disordered $p > p_c(c)$ phases;
\item[$c > 1/4$]
The disordered fixed point is stable for all values of $p$, and no transition occurs.
\end{description}
\end{description}

\begin{description}
\item[Ordered fixed points]
For $O^*\neq 0$, the factor multiplying $O$ in \Cref{order parameter equation for anticonformist} must vanish, which gives the steady state activity as 
\begin{equation}
\label{Ordered_state_s-star}
    s^* = \frac{(1-c)(1-2p)-c}{(1-c)(1-p)}.
\end{equation}
Substituting this $s^*$ into the steady state condition $\dot{s}=0$ in \Cref{activity equation for anticonformist} yields the ordered branch in the standard form 
\begin{equation}
\label{Ordered_state}
    (O^*)^2 = \frac{3(s^*)^2-2s^*}{(1-c)(1-2p)-c}.
\end{equation}
\end{description}

\subsubsection{Order parameter scaling near $p_c(c)$}
In the ordered phase, the non-zero stationary order parameter satisfies \Cref{Ordered_state}. To extract the critical behavior of the order parameter, we define $\delta = p_c(c)-p$ and consider $\delta\ll1$.

Near the critical point $p_c(c)$, the steady state activity behaves as 
\begin{equation}
    s^* \approx \frac{2}{3} + \frac{16}{9}(1-c)\delta.
\end{equation}
Substituting this into the expression of the ordered fixed points in \Cref{Ordered_state}, the order parameter near $p_c(c)$ behaves as 
\begin{equation}
    (O^*)^2 \approx \frac{64}{9}(1-c)\delta.
\end{equation}
Thus, close to the critical point 
\begin{equation}
\label{eq:beta}
    O^* \propto (p_c(c)-p)^{\frac{1}{2}},
\end{equation}
showing that the order parameter exponent is $\beta=1/2$, identical to that of the original model \cite{Biswas_2012}.

\subsubsection{\label{sec:relaxation_timescale}Divergence of the relaxation timescale}
To quantify relaxation towards the disordered fixed point $O^*=0$, we consider a small deviation $O(t)=\epsilon(t)$, with $\epsilon\ll1$. Linearizing the evolution equation for $O(t)$ [\Cref{order parameter equation for anticonformist}] around $O^*=0$, we obtain
\begin{equation}
\label{Linearization of O}
    \frac{d\epsilon}{dt} = \lambda_O \epsilon,
\end{equation}
where $\lambda_O$ is defined in \Cref{eigenvalue of jacobian}.

The solution is $\epsilon(t) = \epsilon(0) \exp(\lambda_0t)$. Hence, the relaxation timescale is defined as 
\begin{equation}
    \tau = \frac{1}{|\lambda_0|}.
\end{equation}
Near the critical point, defining $\delta = p-p_c(c)$, the eigenvalue behaves as 
\begin{equation}
    \lambda_O = -\frac{4}{3}(1-c)\delta.
\end{equation}
Hence, up to a finite prefactor
\begin{equation}
\label{eq:nuz}
    \lambda_0 \propto \delta, \text{ yielding } \tau\propto \delta^{-1}\propto \big(p-p_c(c)\big)^{-1}.
\end{equation}
Therefore, the relaxation time scale diverges at the critical point with an effective exponent $\nu z=1$, exactly the same as in the original model \cite{Biswas_2012}.

The results reported in this section are not surprising as the mean-field model with anticonformists is equivalent to the original BChS model with an effective value of $p$  equal to $c + (1-c)p$. 

\subsection{\label{sec:Mean-field_calculations_for_inflexibles}Mean-field calculations for inflexibles}

\subsubsection{\label{Sec:inflexibles adhered to extreme opinions}Inflexibles adhered to extreme opinions}

We first consider a variant of the original kinetic exchange model in which inflexibles are restricted to agents holding extreme opinions as discussed for intransigents and the quenched case in Section III~B in Reference \onlinecite{Crokidakis_2014a}. We note that Reference \onlinecite{Crokidakis_2014a} provides only numerical results, while here we also present analytical solutions. A fraction $z_+$ of agents with opinion $+1$ and a fraction $z_-$ of agents with opinion $-1$ are inflexibles. Neutral agents with zero opinion are always flexible.

We define three basic probabilities:
\begin{equation}
\label{eq:prob_p+}
    P_+ = \Prob(\mu_{ij}o_j=+1) = (1-p)f_+ + pf_-;
\end{equation}
\begin{equation}
\label{eq:prob_p-}
    P_- = \Prob(\mu_{ij}o_j=-1) = (1-p)f_- + pf_+;
\end{equation}
\begin{equation}
\label{eq:prob_p0}
    P_0 = \Prob(\mu_{ij}o_j=0) = f_0.
\end{equation}

Since only extreme agents can be inflexibles, the transition rates out of $\pm 1$ states are reduced. In particular,
\begin{equation}
\omega_{+1\to 0} = (1-z_+)P_-, \qquad \omega_{-1 \to 0} = (1-z_-)P_+,
\end{equation}
\begin{equation}
    \omega_{+1\to-1}=0,\, \omega_{-1\to+1} = 0,\, \omega_{0\to+1} = P_+,\, \omega_{0 \to -1} = P_-,
\end{equation}
where probabilities $P_+$ and $P_-$ are given by \Cref{eq:prob_p+,eq:prob_p-}.

Using the transition rates listed above, the rate equations for $f_+$ and $f_-$  become
\begin{equation}
\label{f+ equation for inflexibles}
    \frac{df_+(t)}{dt} = P_+f_0 - (1-z_+)P_-f_+,
\end{equation}
\begin{equation}
\label{f- equation for inflexibles}
    \frac{df_-(t)}{dt} = P_-f_0 - (1-z_-)P_+f_-.
\end{equation}
Now, using 
\begin{equation}
\begin{split}
    P_-f_+-P_+f_- = psO,\\
    P_-f_++P_+f_- = \left[\frac{s^2}{2}-\frac{(1-2p)}{2}O^2\right],
\end{split}
\end{equation}
\begin{equation}
 P_+-P_- = (1-2p)O,\quad P_++P_- = s,
\end{equation}
\begin{equation}
 z_+P_-f_+ = \frac{z_+s^2}{4}+\frac{z_+psO}{2}+\frac{z_+O^2}{2}\left(p-\frac{1}{2}\right),
\end{equation}
\begin{equation}
    z_-P_+f_- =  \frac{z_-s^2}{4}-\frac{z_-psO}{2}+\frac{z_-O^2}{2}\left(p-\frac{1}{2}\right)
\end{equation}
and introducing the ratio $\rho = z_+/z_-$ and defining the total fraction of inflexibles as $z=z_+ + z_-$, we obtain
\begin{equation}
     \quad z_+-z_- = \left(\frac{\rho-1}{\rho+1}\right)z.
\end{equation}
By subtracting and adding \Cref{f+ equation for inflexibles,f- equation for inflexibles}, one obtains the  evolution equations for the order parameter $O(t)$ and the activity $s(t)$ as
\begin{equation}
\label{O(t) equation for inflexibles}
\begin{split}
 \frac{dO(t)}{dt}  = (1-s)(1-2p)O -psO +  \left(\frac{\rho-1}{\rho+1}\right)\frac{zs^2}{4}\\ +\frac{zpsO}{2} + \left(\frac{2p-1}{4}\right)\left(\frac{\rho-1}{\rho+1}\right)zO^2,
 \end{split}
\end{equation}
\begin{equation}
\label{s(t) equation for inflexibles}
\begin{split}
    \frac{ds(t)}{dt} = s\left[1+\left(\frac{z}{4}-\frac{3}{2}\right)s\right] + \frac{pz}{2}\left(\frac{\rho-1}{\rho+1}\right)sO\\
    +\frac{(1-2p)}{2}\left(1-\frac{z}{2}\right)O^2.
\end{split}
\end{equation}

In the following we discuss three distinct cases:
\begin{description}
\item[Case I: Symmetric inflexibles $(\rho = 1)$] 
    When $\rho=1$, we have $z_+=z_-$ and therefore $z_+-z_-=0$. As a result, \Cref{O(t) equation for inflexibles,s(t) equation for inflexibles} reduce to
    \begin{equation}
    \label{O equation for symmetric inflexible}
        \frac{dO(t)}{dt} = \left[(1-s)(1-2p)-\left(1-\frac{z}{2}\right)ps\right]O,
    \end{equation}
    \begin{equation}
    \label{s equation for symmetric inflexible}
        \frac{ds(t)}{dt} = s\left[1+\left(\frac{z}{4}-\frac{3}{2}\right)s\right] +\frac{(1-2p)}{2}\left(1-\frac{z}{2}\right)O^2.
    \end{equation}
    One can find the disordered fixed point by setting $O^*=0$ in \Cref{s equation for symmetric inflexible}, which yields the solution $s^* = 4/(6-z)$. The physically relevant disordered fixed point is therefore $[O^*,s^*] = [0,4/(6-z)]$. Linearizing \Cref{O equation for symmetric inflexible} around $[O^*,s^*] = [0,4/(6-z)]$, the eigenvalue controlling the order parameter direction is
    \begin{equation}
        \lambda_O = \left(\frac{2-z}{6-z}\right)(1-4p).
    \end{equation}
    The critical point is obtained from $\lambda_O = 0$, \textit{i.e.},
    \begin{equation}
    \label{eq:pc=1_4}
        p_c = \frac{1}{4},
    \end{equation}
    which is exactly same as in the original model.

\item[Case II: Positive imbalance ($\rho>1$)] 
    Now we consider a small deviation from the symmetric case by introducing a positive imbalance between the two types of inflexibles. We set
    \begin{equation}
        \rho = 1+\epsilon, \qquad  0<\epsilon\ll 1,
    \end{equation}
    so that $z_+-z_->0$.

    Substituting this into \Cref{O(t) equation for inflexibles} and evaluating it at $O^* = 0$, we obtain, to linear order in $\epsilon$,
    \begin{equation}
        \frac{dO(t)}{dt}\Big|_{O^*=0} = \frac{z(s^*)^2}{8}\epsilon.
    \end{equation}
    This contribution is strictly positive for any finite inflexible fraction $z>0$. Hence, irrespective of how small the imbalance $\epsilon$ is, the disordered state $O^*=0$  is no longer a fixed point of the dynamics. Therefore, the notion of the disordered phase ceases to exist once $\rho$ deviates slightly from unity along the positive side. The order parameter remains finite for all values of $p$, and its magnitude increases with increasing inflexible fraction. So, a positive inflexible imbalance acts as a positive bias leading to the suppression of order-disorder phase transition.

\item[Case III: Negative imbalance ($\rho<1$)]
    An analogous conclusion follows for a negative imbalance. Setting
    \begin{equation}
     \rho = 1-\epsilon, \qquad 0<\epsilon\ll 1
    \end{equation}
    we have $z_+-z_-<0$ and 
    \begin{equation}
        \frac{dO(t)}{dt}\Big|_{O^*=0} = -\frac{z(s^*)^2}{8}\epsilon,
    \end{equation}
    yielding the suppression of the order-disorder transition as for Case II.
\end{description}

\subsubsection{\label{Sec:inflexibles adhered to specific opinion}Inflexibles adhered to specific opinion}

We now consider another version of inflexible, where intransigence is restricted to actors holding a specific opinion state, as discussed the intransigents and the quenched case in Section III~C in Reference~\onlinecite{Crokidakis_2014a}. Here, however, we analyze the annealed mean-field version of this model.

\begin{description}
\item[Inflexibles restricted to opinion $+1$]
    We first take a fraction $z_+=z$ of agents with opinion $+1$ to be inflexibles,
    while $z_-=z_0=0$. In terms of the imbalance parameter, this corresponds to the extreme limit $\rho = z_+/z_- \to \infty$, which is the boundary case of a positive imbalance. Based on the previous analysis, we therefore expect behavior analogous to that induced by a positive bias.

    In this case, only the transition $\omega_{+1\to0}$ is affected, being suppressed by a factor of $(1-z)$. The resulting evolution equation for the order parameter becomes,
    \begin{equation}
    \begin{split}
        \frac{dO(t)}{dt} = \left[(1-s)(1-2p)+\frac{ps}{2}(z-2)\right]O\\+\frac{zs^2}{4}+\frac{z(2p-1)}{4}O^2.
    \end{split}
    \end{equation}
    Evaluating this equation at $O^*=0$ yields 
    \begin{equation}
     \frac{dO(t)}{dt}\Big|_{O^*=0} = \frac{z(s^*)^2}{4}.
    \end{equation}
    Since the contribution is strictly positive for any finite inflexible fraction $z>0$, the disordered fixed point $O^*=0$ ceases to exist. Consequently, the order-disorder phase transition is suppressed, as inflexibles restricted to the $+1$ state act as an effective positive bias.

\item[Inflexibles restricted to opinion $-1$] 
    Next, we consider the opposite extreme case with $z_-=z$ and $z_+=z_0=0$, corresponding to $\rho=0$. This represents the boundary case of a negative imbalance, and we therefore expect behavior analogous to that induced by a negative bias.

    Here, only the transition $\omega_{-1\to0}$ is modified. The resulting order parameter evolution equation reads 
    \begin{equation}
    \begin{split}
        \frac{dO(t)}{dt} = \left[(1-s)(1-2p)+\frac{ps}{2}(z-2)\right]O\\-\frac{zs^2}{4}-\frac{z(2p-1)}{4}O^2.
    \end{split}
    \end{equation}
    Evaluating this equation at $O^*=0$ yields 
    \begin{equation}
     \frac{dO(t)}{dt}\Big|_{O^*=0} = -\frac{z(s^*)^2}{4}.
    \end{equation}
    and similarly to the previous case such inflexibles act as an effective negative bias.

\item[Inflexibles restricted to neutral  opinion $0$] 
    We now consider the case where inflexibles are restricted exclusively to the neutral state i.e. $z_+=z_-=0$ and $z_0 = z$. Since, there is no imbalance between the extreme opinions, this corresponds to a zero bias situation, and therefore one expects the system to behave similarly to the unbiased case, with an order-disorder phase transition at a finite critical point.

    In this case, only the transitions $0\to+1$ and $0\to-1$ are affected by the presence of inflexibles. The resulting mean-field evolution equations for the order parameter $O$ and the activity $s$ reads,
    \begin{equation}
    \label{O equation for zero inflexible}
        \frac{dO(t)}{dt} = \left[(1-s)(1-2p)(1-z)-ps\right]O,
    \end{equation}
    \begin{equation}
    \label{s equation for zero inflexible}
        \frac{ds(t)}{dt}= s(1-s)(1-z)-\frac{s^2}{2}+\frac{(1-2p)}{2}O^2.
    \end{equation}
    Setting $O^*=0$ in \Cref{s equation for zero inflexible}, one finds the physically relevant disordered fixed point 
    \begin{equation}
        (O^*,s^*) = \left(0,\frac{2(1-z)}{3-2z}\right).
    \end{equation}
    Linearizing \Cref{O equation for zero inflexible} around this disordered fixed point yields the eigenvalue associated with the order-parameter direction,
    \begin{equation}
        \lambda_O = \left(\frac{1-z}{3-2z}\right)(1-4p).
    \end{equation}
    The critical point is therefore given by \Cref{eq:pc=1_4},
    which is exactly the same as in the original model and coincides with the result reported  in Section~C of Reference \onlinecite{Crokidakis_2014a} for the quenched version of neutral intransigents.
\end{description}

\subsubsection{\label{Sec:inflexibles chosen independently of their initial opinion}Inflexibles chosen independently of their initial opinion}

We now turn to another version of inflexibles, where a fraction $z$ of agents is chosen uniformly at random, independently of their initial opinion, as discussed in Section III~A of Reference \onlinecite{Crokidakis_2014a} for intransigent and the quenched case. Here, we analyze the corresponding annealed mean-field version.

Since inflexibles never update their opinions, all transition rates involving opinion updates are suppressed by a factor $(1-z)$. With this modification, the mean-field evolution equation for the order parameter becomes 
\begin{equation}
\label{O equation for random inflexible}
    \frac{dO(t)}{dt} = (1-z)\left[(1-s)(1-2p)-ps\right]O,
\end{equation}
while the activity evolves according to
\begin{equation}
\label{s equation for random inflexible}
    \frac{ds(t)}{dt} = (1-z)\left[s(1-s)-\frac{s^2}{2}+\frac{(1-2p)}{2}O^2\right].
\end{equation}
Setting $O^*=0$, \Cref{s equation for random inflexible} yields the same disordered fixed point as in the original model, $(O^*,s^*) = (0,2/3)$, independent of the inflexible fraction $z$.

Linearizing \Cref{O equation for random inflexible} around this disordered fixed point gives the eigenvalue 
\begin{equation}
    \lambda_O = \frac{1-z}{3}(1-4p).
\end{equation}
The critical point is therefore again given by \Cref{eq:pc=1_4},
which is unchanged by the presence of annealed inflexibles. This is in clear contrast to the quenched case reported in Equation (11) of Section III A of Reference \onlinecite{Crokidakis_2014a}, where the critical point shifts with the inflexible fraction.

Substituting $O^*\neq 0$ and the stationary activity $s^*=2/3$ into \Cref{s equation for random inflexible} yields the ordered fixed points for $p<1/4$
\begin{equation}
\label{eq:fixed_point_inflexibles_independent_on_initial_opinion}
    (O^*)^2 = \frac{1-4p}{(1-p)^2},
\end{equation}
which is independent of the inflexible fraction $z$. Hence in the annealed case, inflexibles merely rescale the timescale of the dynamics but do not affect the location of the critical point or the stationary order parameter.

\subsection{\label{sec:Monte_Carlo_simulations}Monte Carlo simulations}

\begin{figure}[bp]
(a) 

\includegraphics[width=0.90\columnwidth]{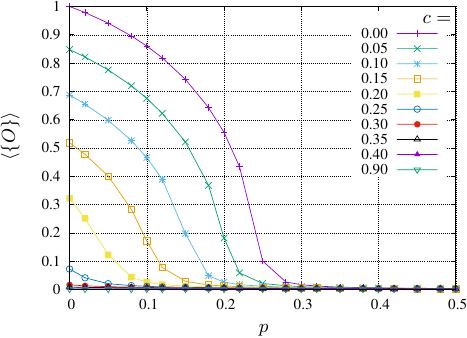}

(b) 

\includegraphics[width=0.90\columnwidth]{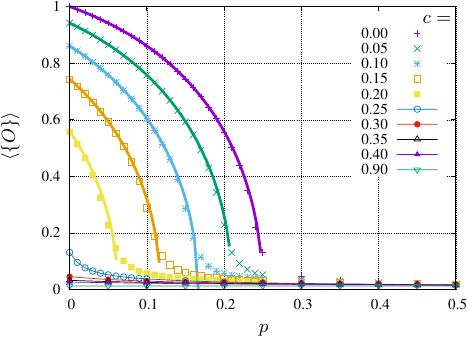}
\caption{\label{fig:rev1_o_vs_p_c_disc}Mean order parameter $\langle\{O\}\rangle$ for various fractions $c$ of anticonformists with discrete opinions.
The (a) quenched or (b) annealed disorder is assumed.
In panel (b) the thick solid lines (for $c\le 0.2$) represent the closed form expression of the order parameter given in \Cref{Ordered_state_s-star,Ordered_state} while the points denote the Monte Carlo simulation results.
$N=1024$, $T=10^3$, $R=10^3$, $\tau=0.2T$.}
\end{figure}

The simulation is performed on the fully connected graph with $N$ nodes.
In every elementary time step, the actor $i$ is randomly selected and her/his partner $j$ for interaction with $\mu_{ij}$ according to the rules of the model is also randomly selected.
A single Monte Carlo step is completed after $N$ such elementary time steps.
Symbols $\{\cdots\}$ and $\langle\cdots\rangle$ stand for a time average over the last $\tau$ time steps and average over $R$ independent simulations, respectively.
All simulations are performed up to a maximum system size $N=1024$ and averaged over $R=10^3$ independent simulations.
For a single simulation, the results are gathered from the last $\tau=0.2T$ time steps, where  $T=10^3$ is the total number of Monte Carlo steps allowed.
 Quenched and annealed disorders have been considered.
For the annealed disorder, every selected actor decides to behave as anticonformist (inflexible) with probability $c$ ($z$).
For the quenched disorder, the identities of the anticonformists (inflexibles) in a single simulation are fixed and their fraction is $c$ ($z$).

To obtain the phase diagram in the $(c,p)$ space for anticonformists---besides the case of discrete opinions and annealed disorder, where the phase diagram is given analytically by \Cref{eq:pc_vs_c}---we compute the fourth-order Binder cumulant (see Reference~\onlinecite[p.~78]{Guide_to_Monte_Carlo_Simulations_2009}) of the order parameter $O$
\begin{equation}
\label{eq:Binder}
U= 1-\frac{\{ O^4\}}{3\{ O^2\}^2}.
\end{equation}
$U(p)$ plotted for various system sizes $N$ has a single common crossing-point that predicts the critical value of $p_c$ for fixed fraction $c$ of anticonformists. 
For getting smooth curves of $U(p)$ we increase the simulation time to $T=10^4$.

\subsubsection{Discrete opinions}

For discrete opinions, we have $o_i\in\{-1,0,+1\}$ and $\mu_{ij}=\pm 1$.
The fraction of negative values of $\mu_{ij}$ is $p$.

\Cref{fig:rev1_o_vs_p_c_disc}(a) presents the mean order parameter $\langle\{O\}\rangle$ for various fractions $c$ of anticonformists with discrete opinions.

\begin{figure}[htbp]
(a)

\includegraphics[width=0.90\columnwidth]{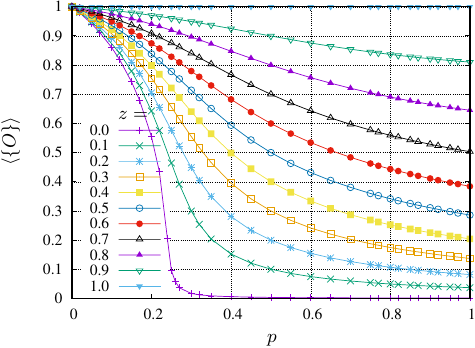}

(b)

\includegraphics[width=0.90\columnwidth]{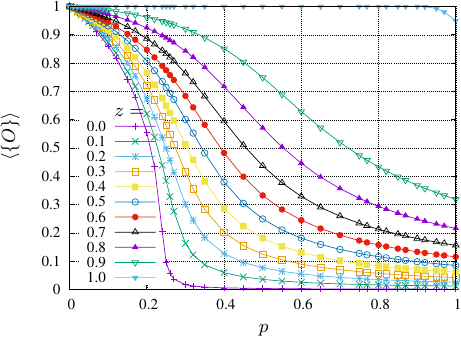}
\caption{\label{fig:o_vs_p_z_disc}Mean order parameter $\langle\{O\}\rangle$ for various fractions $z$ of inflexibles with discrete opinions.
The inflexibles are adhered to opinion $+1$.
The (a) quenched or (b) annealed disorder is assumed.
$N=1024$, $T=10^3$, $R=10^3$, $\tau=0.2T$.}
\end{figure}

\Cref{fig:o_vs_p_z_disc} shows the mean order parameter $\langle\{ O\}\rangle$ for various fractions $z$ of inflexibles with discrete opinions.
The quenched [\Cref{fig:o_vs_p_z_disc}(a)] or annealed [\Cref{fig:o_vs_p_z_disc}(b)] disorder is assumed.
For the quenched disorder fraction of $z$ actors, marked as inflexibles, it has a fixed opinion $+1$.
For the annealed disorder, the selected actor, with probability $z$ takes the opinion $+1$ instead of following the rule \eqref{eq:model}.

\begin{figure}
(a)

\includegraphics[width=0.90\columnwidth]{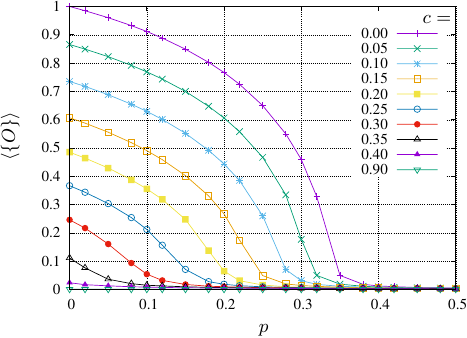}

(b)

\includegraphics[width=0.90\columnwidth]{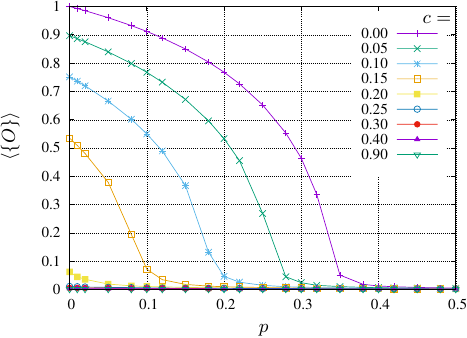}
\caption{\label{fig:rev1_o_vs_p_c_cont}Mean order parameter $\langle\{O\}\rangle$ for various fractions $c$ of anticonformists with continuous opinions.
The (a) quenched or (b) annealed disorder is assumed.
$N=1024$, $T=10^3$, $R=10^3$, $\tau=0.2T$.}
\end{figure}

\subsubsection{Continuous opinions}

For continuous opinions, we have $o_i\in[-1;+1]$ and $\mu_{ij}=\pm 1$.
The fraction $c$ of agents is considered anticonformist and $p$ is a fraction of negative values of $\mu_{ij}$.

\Cref{fig:rev1_o_vs_p_c_cont} shows the order parameter $\langle\{O\}\rangle$ for various fractions $c$ of anticonformists with continuous opinions.

\begin{figure}
\includegraphics[width=0.90\columnwidth]{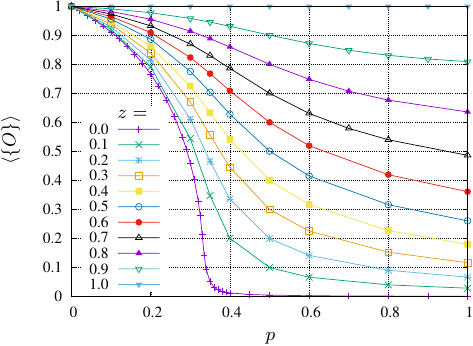}
\caption{\label{fig:o_vs_p_z_cont}Mean order parameter $\langle\{O\}\rangle$ for various fractions $z$ of inflexibles for continuous opinions. 
The inflexibles are adhered to opinion $+1$.
Only the quenched disorder may be considered.
$N=1024$, $T=10^3$, $R=10^3$, $\tau=0.2T$.}
\end{figure}

In \Cref{fig:o_vs_p_z_cont} the order parameter $\langle\{O\}\rangle$ is presented for various fractions $z$ of inflexibles with continuous opinions.
Inflexibles are adhered to opinion $+1$.
For continuous opinions only quenched disorder is available as the chance of selecting actors with given opinion (for instance $o_i=\pm 1$) becomes negligible.

\section{\label{sec:discussion}Discussions}

\begin{figure}[htbp]
(a) 

\includegraphics[width=0.60\columnwidth]{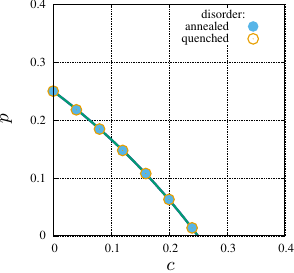}

(b) 

\includegraphics[width=0.60\columnwidth]{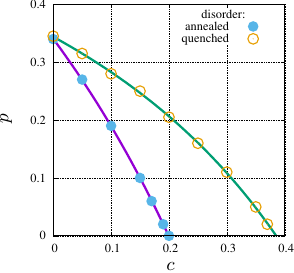}
\caption{\label{fig:phase_diagrams}Phase diagram on $(p,c)$ plane for anticonformists and
(a) discrete opinions; solid line shows \Cref{eq:pc_vs_c}, 
(b) continuous opinions; solid lines show fits \eqref{eq:phase_diagram_c_c_a} and \eqref{eq:phase_diagram_c_c_q}.} 
\end{figure}

For anticonformists, the mean-field approach leads to conclusions that an ordered phase is available when the fraction of anticonformists is $c<1/4$ [see \Cref{eq:pc_vs_c}].
This is also clearly visible in \Cref{fig:rev1_o_vs_p_c_disc}.
For the version with discrete opinions and annealed disorder, the results of Monte Carlo simulations [see \Cref{fig:rev1_o_vs_p_c_disc}(b)] perfectly suit the results of mean-field analysis [\Cref{Ordered_state}] as presented in \Cref{fig:rev1_o_vs_p_c_disc}(b).
The phase diagram in the $(p,c)$ plane for this case is given by \Cref{eq:pc_vs_c} and presented in \Cref{fig:phase_diagrams}(a).

\begin{figure*}[htbp]
\hspace{3cm}(a) \hfill (b) \hfill (c)\hspace{2cm}

      \includegraphics[width=.30\textwidth]{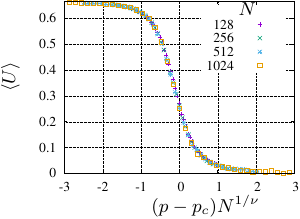}
\hfill\includegraphics[width=.30\textwidth]{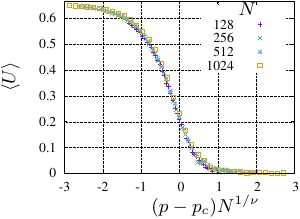}
\hfill\includegraphics[width=.30\textwidth]{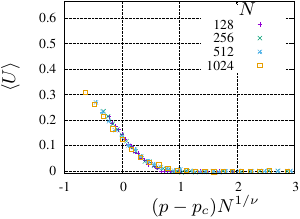}\\

\hspace{3cm}(d) \hfill (e) \hfill (f)\hspace{2cm}

      \includegraphics[width=.30\textwidth]{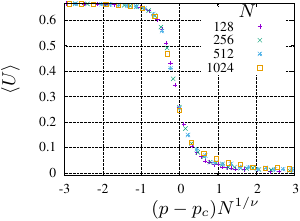}
\hfill\includegraphics[width=.30\textwidth]{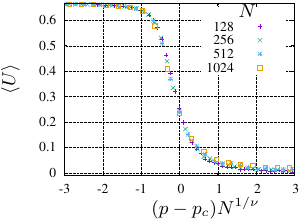}
\hfill\includegraphics[width=.30\textwidth]{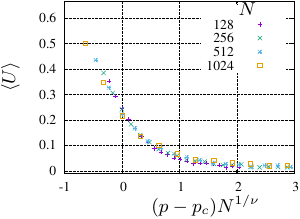}\\

\hspace{3cm}(g) \hfill (h) \hfill (i)\hspace{2cm}

      \includegraphics[width=.30\textwidth]{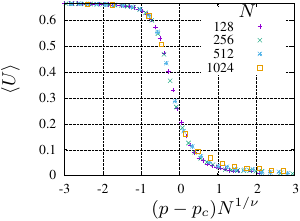}
\hfill\includegraphics[width=.30\textwidth]{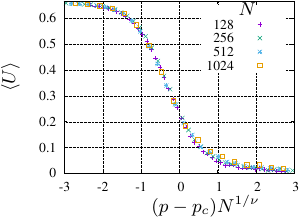}
\hfill\includegraphics[width=.30\textwidth]{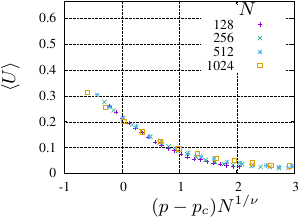}

\caption{\label{fig:Binder}Examples of Binder cumulants $\langle U\rangle$ \eqref{eq:Binder} for various system sizes $N$ and various fractions $c$ of anticonformists as dependent on $(p-p_c)N^{1/\nu}$ with $\nu=2$.
The $p_c$ values are roughly estimated by an inspection these plots before rescaling the $x$-axis.
(a)--(c) discrete opinions and annealed disorder:
(a) $c=0.05$,
(b) $c=0.15$,
(c) $c=0.24$;
(d)--(f) continuous opinions and annealed disorder:
(d) $c=0.05$,
(e) $c=0.1$,
(f) $c=0.19$;
(g)--(i) continuous opinions and quenched disorder:
(g) $c=0.05$,
(h) $c=0.2$,
(i) $c=0.37$.
$T=10^4$, $\tau=0.2T$, $R=10^3$.} 
\end{figure*}

For construction of the phase diagram in the $(p,c)$ plane for discrete opinions with quenched disorder and for continuous opinions one must rely solely on Monte Carlo simulation.
To that end, we compute the Binder cumulants $U(p)$ \eqref{eq:Binder}.
Few examples of the Binder cumulants for $N=128$, 256, 512 and 1024 as dependent on $(p-p_c)N^{1/\nu}$ with $\nu=2$ are presented in \Cref{fig:Binder}.

For discrete opinions and quenched disorder the results of Monte Carlo simulation follow again mean-field predictions \eqref{eq:pc_vs_c} as presented in \Cref{fig:phase_diagrams}(a).

For continuous opinions and for both annealed and quenched disorder, the phase boundaries in the $(c,p)$ plane are shown in \Cref{fig:phase_diagrams}(b).
Motivated by the form of the phase boundary, Eq.~\eqref{eq:pc_vs_c}, which has been obtained analytically, we attempt a similar reasonable fitting function here. 
For the annealed case, the method of least-squares leads to
\begin{equation}
\label{eq:phase_diagram_c_c_a}
p_c(c)=\frac{1-a_1c}{a_2-a_3c}
\end{equation}
with 
$a_1=5.0118(82)$,
$a_2=2.9363(76)$,
$a_3=3.05(14)$.

The critical value $p_c \approx 0.34$ in the absence of anticonformists (for $c=0$) has already been reported in the original article \cite{Biswas_2012}.
For $c>0.2$, independent of the values of $p$, only a disordered phase is observed.
For continuous opinions and quenched disorder, the ordered phase is available for $c<0.38$ and fit \eqref{eq:phase_diagram_c_c_a} gives
\begin{equation}
\label{eq:phase_diagram_c_c_q}
a_1=2.595(10),\,
a_2=2.907(14),\,
a_3=2.89(13).
\end{equation}

\begin{figure*}[htbp]
\hspace{4cm}(a) \hfill (b) \hspace{3cm}

\includegraphics[width=0.90\columnwidth]{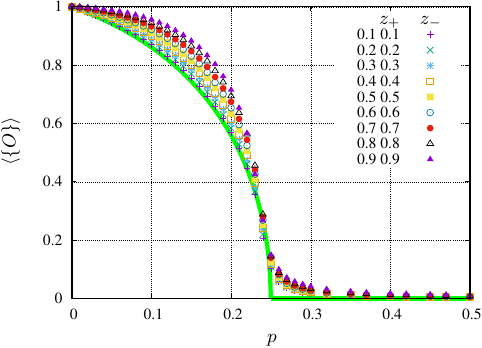}
\hfill\includegraphics[width=0.90\columnwidth]{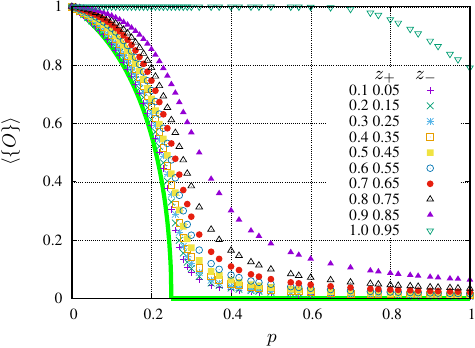}
\caption{\label{fig:inflexibles_o_vs_p_annealed_cases12}Mean order parameter $\langle\{ O\}\rangle$ for various probabilities $z_+$ and $z_-$ of inflexibles adhered to extreme opinions ($+1$ and $-1$, respectively).
(a) Case I in \Cref{Sec:inflexibles adhered to extreme opinions} ($\rho=1$, $z_+=z_-$) and (b) Case II in \Cref{Sec:inflexibles adhered to extreme opinions} ($\rho>1$, $z_+>z_-$). 
Discrete opinions and annealed disorder are assumed.
The solid line represents the closed form expression of the order parameter given in \Cref{eq:fixed_point_inflexibles_independent_on_initial_opinion} while the points denote the simulation results.
All simulations were done for the system size $N = 1024$}
\end{figure*}

\begin{figure}[htbp]
\includegraphics[width=0.90\columnwidth]{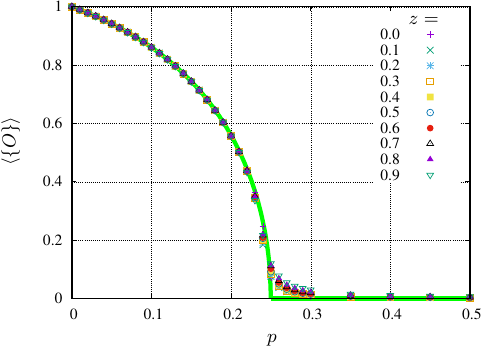}
\caption{\label{fig:inflexibles_o_vs_p_annealed_v3}Mean order parameter $\langle\{ O\}\rangle$ for various fractions $z$ of inflexibles chosen independently of their initial opinion (\Cref{Sec:inflexibles chosen independently of their initial opinion}).
Discrete opinions and annealed disorder are assumed.
The solid line represents the closed form expression of the order parameter given in \Cref{eq:fixed_point_inflexibles_independent_on_initial_opinion} while the points denote the simulation results.
All simulations were done for the system size $N = 1024$.}
\end{figure}

\begin{figure*}[htbp]
\hspace{2cm}(a) \hfill (b) \hfill (c)\hspace{2cm}

      \includegraphics[width=.31\linewidth]{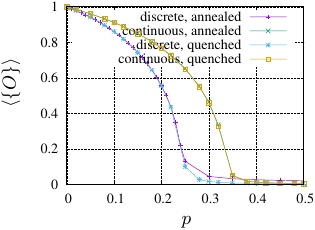}
\hfill\includegraphics[width=.31\linewidth]{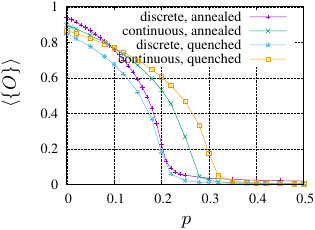}
\hfill\includegraphics[width=.31\linewidth]{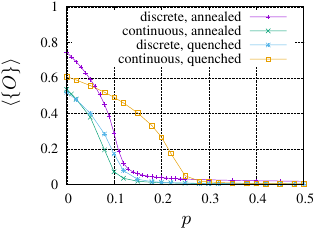}
\caption{\label{fig:comparing_c}Mean order parameter $\langle\{O\}\rangle$ for fractions (a) $c=0.0$---original model formulation, (b) $c=0.05$ and (c) $c=0.15$ of anticonformists and discrete/continuous opinions and the annealed/quenched disorder.
The anticonformist $i$ has alignment index $\mu_{ij}=-1$ for every actor $j$ and occupy fixed fraction $c$ of position on network (quenched disorder) or behave as anticonformists taking $\mu_{ij}=-1$ with probability $c$ (annealed disorder).
$N=1024$, $T=10^3$, $R=10^3$, $\tau=0.2T$.}
\end{figure*}

\begin{figure*}[htbp]
\hspace{2cm}(a) \hfill (b) \hfill (c)\hspace{2cm}

      \includegraphics[width=.31\linewidth]{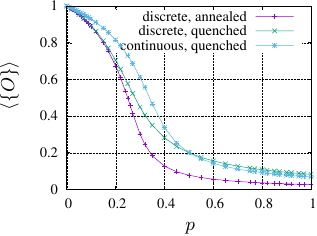}
\hfill\includegraphics[width=.31\linewidth]{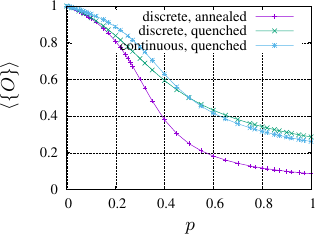}
\hfill\includegraphics[width=.31\linewidth]{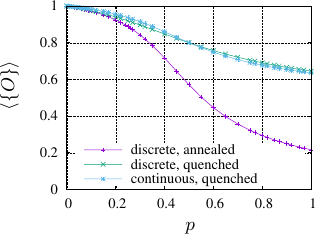}
\caption{\label{fig:comparing_z}Mean order parameter $\langle\{O\}\rangle$ for fractions (a) $z=0.2$, (b) $z=0.5$ and (c) $z=0.8$ of inflexibles and discrete/continuous opinions and the quenched/annealed disorder.
The inflexibles keep opinion ($+1$) with probability $z$ (for annealed disorder) or have opinion ($+1$) and fixed position on network (for quenched disorder).
In the latter the fraction $z$ of nodes is marked as occupied by inflexibles.
$N=1024$, $T=10^3$, $R=10^3$, $\tau=0.2T$.}
\end{figure*}

For inflexibles adhered to extreme opinions (\Cref{Sec:inflexibles adhered to extreme opinions}), the symmetric point $\rho=1$ [with the fractions of inflexibles adhered to opinion $+1$ equal to the fractions of inflexibles adhered to opinion $-1$, see \Cref{fig:inflexibles_o_vs_p_annealed_cases12}(a)] is special. It is the only value of $\rho$ for which the order parameter equation is invariant under $O\to -O$ and admits a disordered fixed point. 
Any infinitesimal deviation from $\rho=1$ [as presented in \Cref{fig:inflexibles_o_vs_p_annealed_cases12}(b) for $z_+>z_-$] destroys this symmetry and suppresses the phase transition. In this sense, $\rho=1$ is a structurally unstable point protected by symmetry that separates unbiased dynamics from biased ones.

Clearly, for the case with both opinions and the interactions taking continuous values, the phase boundaries are different for quenched and annealed disorder. To check whether this difference arises because of taking  both opinions and interactions  continuous, we also compared results (not shown)  for the two types of disorder when the opinions are continuous but $\mu_{ij}$ are discrete. In this case, however, the results are similar quantitatively to the case with discrete opinions and identical for the two types of disorder. So one can conclude that only when interactions are continuous, the continuous opinion case results are determined by the nature of the disorder.

The introduction of inflexibles (\Cref{Sec:inflexibles adhered to specific opinion}) adhered to a specific opinion (for example, $+1$) leads to a subsequent increase in the order parameter with $z$ (up to $O\equiv 1$, independent of $p$, for $z=1$). Subsequently, the order-disorder phase transition is suppressed (see \Cref{fig:o_vs_p_z_disc}). 

For inflexibles chosen independently of their initial opinion (\Cref{Sec:inflexibles chosen independently of their initial opinion}) the mean order parameter $\langle\{O\}\rangle$ is independent of their fraction $z$ as predicted by mean-field calculations \eqref{eq:fixed_point_inflexibles_independent_on_initial_opinion} and shown in \Cref{fig:inflexibles_o_vs_p_annealed_v3}.

The results of the simulations are qualitatively similar for discrete or continuous opinions (\Cref{fig:rev1_o_vs_p_c_disc,fig:rev1_o_vs_p_c_cont}), (\Cref{fig:o_vs_p_z_disc,fig:o_vs_p_z_cont}), indicating that the exact nature of the opinion values does not matter.

\Cref{fig:comparing_c} shows the mean order parameter $\langle\{O\}\rangle$ as a function of $p$ with anticonformist fractions $c=0.0$ [\Cref{fig:comparing_c}(a)---original model formulation], $c=0.05$ [\Cref{fig:comparing_c}(b)] and $c=0.15$ [\Cref{fig:comparing_c}(c)] for discrete/continuous opinions and quenched/annealed disorder.
In all cases a systematic out-performance of the mean order parameter for the continuous space of opinions is observed, which is consistent with the  original higher critical value of $p_c$---being $p_c\approx 0.34$ for the continuous version of the opinion space and $p_c= 1/4$ for the discrete version of the opinion space \cite{Biswas_2012}.

\Cref{fig:comparing_z} shows the mean order parameter $\langle\{O\}\rangle$ for fractions $z=0.2$ [\Cref{fig:comparing_z}(a)], $z=0.5$ [\Cref{fig:comparing_z}(b)] and $z=0.8$ [\Cref{fig:comparing_z}(c)] of inflexibles and discrete/continuous opinions and the quenched/annealed disorder.
For continuous opinions, only quenched disorder can be simulated, as the probability of random selection of the actor $i$ with the preselected opinion (neutral $o_i=0$, or extreme $o_i=\pm 1$, or for that matter, any specific value) becomes negligible.
For quenched disorder, the results of the Monte Carlo simulations for discrete opinions do not differ much from those for continuous ones.
But  $\langle\{O\}\rangle$ for discrete opinions and annealed disorder is systematically lower than $\langle\{O\}\rangle$ for discrete opinions and quenched disorder. 
This difference is better visible for higher values of $p$ ($p>0.2$) and particularly for high values of $z$. 
In the version with annealed disorder, every actor can suddenly leave his/her opinion (for example $+1$), while for quenched disorder, inflexibles with extreme opinions never change, and thus keeping higher values of $\langle\{O\}\rangle$ is easier.

From a finite-size scaling analysis of the Binder cumulant \eqref{eq:Binder} and the fluctuations of the order parameter, the critical exponents $\nu$ and $\gamma$ can be obtained (not shown)
for the model with anticonformists, for which an order-disorder transition is obtained. Together with the analytically determined exponents $\beta$ [\Cref{eq:beta}] and $\nu z$ [\Cref{eq:nuz}], these results confirm that the critical behavior belongs to the same mean-field Ising universality class as the original BChS model \cite{Biswas_2012}. 

From a sociological perspective, the results of calculations and simulations clearly show the influence of anticonformists and inflexibles on the formation of opinion in society: the presence of anticonformists decrease the order parameter, while the presence of inflexibles increase the order parameter.
Since the \citeauthor{moscovici_1969} experiment \cite{moscovici_1969}, we realize that minorities can influence a majority (see Reference \onlinecite{Moscovici_1976} for a summary and References \onlinecite{Siedlecki_2016,Krueger_2017,1906.06094,Galam_2007366} for some recent studies).
This influence is significant only when they behave in certain ways.
Anticonformists oppose the majority `on principle', while inflexibles keep their opinions independent of group opinion.
Such a consistent behavior is one crucial in influencing the group by minorities, and the simulations presented here support this statement.
We must also admit that simulations with quenched disorder are better motivated from a sociological perspective---then actors are anticonformists (or inflexibles) and not only behave as anticonformists (or inflexibles) with probability $c$ (or $z$).

What is the reason for not getting an order-disorder transition in the presence of inflexibles except for some special cases? We have shown that an unbalanced ratio of inflexibles belonging to opinions $\pm 1$ in the discrete case favors a certain extreme opinion, resulting in a general drift in opinion.
 In the continuous opinion case with quenched disorder, in presence of  anticonformists on the other hand, the value of $p_c$ increases. These two cases may have a connection in the sense that in the continuous version with anticonformists, the updated opinions may still remain of the same sign even with $\mu_{ij} $ negative, which results in having a approximate inflexible behavior. Of course, such inflexibles may favor either a positive or negative opinion and also such behavior may not always occur, which is why one still gets a transition, albeit at a larger value of $p_c$.

In conclusion, in this paper, the BChS model is extended by introducing anticonformists and inflexibles.
With  Monte Carlo simulations and mean-field calculations, we show the influence of fraction $c$ of anticonformists and fraction $z$ of inflexibles on the order parameter $O$, which captures net polarization.
With the mean-field calculations, we identified fractions $c$ of anticonformists that lead to ordered phases.
For Monte Carlo simulations, discrete/continuous opinions were introduced, and annealed/quenched disorder was considered.
For discrete opinions and annealed disorder, the Monte Carlo simulations perfectly match the mean-field calculations both, for anticonformists and inflexibles.
The qualitative results of introducing anticonformists (inflexibles) in various ways (discrete/continuous opinions and annealed/quenched disorder) are roughly the same. 
However, for the BChS model extended by inflexibles adhered to opinion $+1$, we can see a systematic shift of the mean order parameter to its higher values for quenched disorder than for the variant with annealed disorder. 

Further studies may include mixing anticonformists with inflexibles or checking spatial correlation for various types of actors for quenched disorder, where the positions of anticonformists and inflexibles are fixed on the graph. Perhaps one can look at the persistence behavior of the actors. Finally, also the underlying complete graph may be replaced by a closer to real-world social network.

\begin{acknowledgments}
KM thanks PS for her hospitality in the University of Calcutta, Kolkata.
KM visit to Kolkata was supported by `Excellence initiative---research university' program for the AGH University of Krakow.
AP acknowledges University Grant Commission (UGC), Government of India, for financial support (NTA Reference No. 241610061476). 
PS is grateful to RUSA 2.0 financial grant (Ministry of Education, Government of India).
\end{acknowledgments}


%
\end{document}